\documentclass[11pt,a4paper]{article}
\usepackage{amsfonts}
\usepackage{color}
\usepackage{amssymb}
\usepackage{amsmath}
\usepackage{graphicx,color}
\usepackage{amsthm}
\usepackage[latin9]{inputenc} 
\usepackage{enumerate}
\usepackage{siunitx}
\usepackage{float}
\newtheorem*{lemma*}{Lemma}

\numberwithin{equation}{section}
\date{}

\newcommand{\be}{\begin{equation}}
\newcommand{\ee}{\end{equation}}

\begin{document}

\title{Band spectra of periodic hybrid $\delta$-$\delta^\prime$ structures}

\author{M. Gadella$^1$\footnote{manuelgadella1@gmail.com}, 
J. M. Mateos Guilarte$^2$\footnote{guilarte@usal.es}, 
J. M. Mu\~noz-Casta\~neda$^1$\footnote{jose.munoz.castaneda@uva.es},  \\[1ex]
L. M. Nieto$^1$\footnote{luismiguel.nieto.calzada@uva.es}, and
L. Santamar\'{\i}a-Sanz$^1$\footnote{lucia.santamaria@uva.es}
\\ [2ex]
\footnotesize{\sl $^1$Departamento de F\'{\i}sica Te\'{o}rica, At\'{o}mica y \'{O}ptica and IMUVA,
Univ. de Valladolid, 47011 Valladolid, Spain}
\\
\footnotesize{\sl $^2$Departamento de F\'{\i}sica Fundamental, IUFFyM,
Universidad de Salamanca, Spain}}

\date{\today}

\maketitle

\begin{abstract}
We present a detailed study of a generalised one-dimensional Kronig-Penney model using $\delta\text{-}\delta'$ potentials.  We analyse the band structure and the density of states in two situations. In the first case we consider an infinite array formed by identical $\delta\text{-}\delta'$ potentials standing at the linear lattice nodes. This case will be known throughout the paper as the one-species hybrid Dirac comb. We investigate the consequences of adding the $\delta'$ interaction to the Dirac comb by comparing the band spectra and the density of states of pure Dirac-$\delta$ combs and one-species hybrid Dirac combs. Secondly we study the quantum system that arises when the periodic potential is the one obtained from the superposition of two one-species hybrid Dirac combs displaced one with respect to the other and with different couplings. The latter will be known as the two-species hybrid Dirac comb. One of the most remarkable results is the appearance of a curvature change in the band spectrum when the $\delta'$ couplings are above a critical value.

\end{abstract}

\section{Introduction}

In this paper we perform an analysis of a generalised Kronnig-Penney model built from $\delta\text{-}\delta'$ potentials. The Kronig-Penney model is a well known example of a one-dimensional periodic potential used in solid state physics to describe the motion of an electron in a periodic array of rectangular barriers or wells \cite{kronig-prsa31}. One variation of this model is the so called Dirac comb in which the rectangular barriers/wells become Dirac delta potentials  with positive or negative strength respectively. The Hamiltonian for the Dirac comb is
\begin{equation}\label{1.1}
{\cal H}= -\frac{\hbar^2}{2m}\,\frac{d^2}{dy^2}+V_1(y), \quad \text{where} \quad
V_1(y)=\mu \sum_{n=-\infty}^\infty \delta(y-n y_0),\quad \mu \in \mathbb{R},\quad  y_0>0,
\end{equation}
where both parameters $\mu$ and $y_0$ are fixed.

Dirac delta type potentials are exactly solvable models frequently used to describe quantum systems with very short range interactions which are located around a given point. These two properties make them suitable to obtain many general properties of realistic quantum systems  \cite{textbooks,textbooks2,textbooks3}. Moreover Dirac delta potentials enable the study of the Bose-Einstein condensation in periodic backgrounds \cite{bordagJPA2019}, in a harmonic trap with a tight and deep ``dimple'' potential, modelled by a Dirac delta function \cite{Uncu}, or a nonperturbative study of the entanglement of two directed polymers subject to repulsive interactions given by a Dirac delta function potential \cite{Ferrari}. It is also interesting to use a Dirac comb to investigate the light propagation (transverse electric and magnetic modes as well as omnidirectional polarization modes) in a  one-dimensional relativistic dielectric superlattice \cite{alv-rod-pre99,zur-san-pre99,lin-pre06}. These types of interactions have been used in other contexts such as in studies related with supersymmetry  \cite{ignacio1,ignacio2,ignacio3,ignacio4,guilarte-epjp130}. 

It is of note that although the rigorous definition of the Dirac delta potential in one dimension is well known, reproducing the definition of the Dirac delta potential standing in one point for two and three dimensional spaces is highly non-trivial \cite{textbooks3,jackiw,bordag-prd91,bordag-prd95} and requires the use of the theory of self adjoint extensions to introduce a regularization parameter.

Contact interactions or potentials, also known as zero-range potentials, can be understood as generalisations of the Dirac-$\delta$ potential by means of boundary conditions (see subsection 3.1.1 in Ref. \cite{pistones}). These types of potentials have been largely used in different areas of physics over the past 40 years. Their importance is specially relevant for applications to atomic physics developed in the 80s (see \cite{zrpap} and references therein). New mathematical tools have been introduced in physics in order to define, characterise and classify rigorously contact potentials \cite{ALB1,AK}. After the seminal papers by Berezin and Faddeev \cite{fad} and Kurasov's paper where contact potentials are characterised by certain self adjoint extensions of  the one-dimensional  kinetic operator $K=-d^2/dx^2$ \cite{KUR}, several attempts have been made to explain the physical meaning of the contact potentials that emerge from these extensions \cite{KP,KP1}. More recently there have been papers where, contact potentials have been used to study the effects of resonant tunneling  \cite{ZOL,ZOL1}, to study their properties under the effect of external fields \cite{ZOL2}, and their applications in the study of metamaterials \cite{NIE}. In addition the effects of several contact potential barriers have been studied in \cite{LUN,KOR,EGU,EGU1,gadella-jpa16,Caudrelier}  and extensions to arbitrary dimension have been considered in \cite{MNR} generalising the approach by Jackiw in \cite{jackiw}.  Mathematical properties of potentials decorated with contact interactions have also been the object of recent studies \cite{LAR,ANT,GOL,GOL1,ALB,ET,SPL}. Finally, we would like to mention   a wide range of physical applications that have appeared in the last year \cite{KP2,RBO,CAL,JUAN}.  

Among these possible generalizations of the Dirac delta potential the most obvious to start with is the derivative of the Dirac delta, usually denoted as $\delta'(y)$. In fact, this interaction has already been considered by several authors in the past \cite{textbooks3,SE,FAS,bawa}. In combination with the Dirac delta produces a potential of the form
\begin{equation}\label{1.2}
V_2(y)= \mu\,\delta(y)+\lambda\,\delta'(y),
\end{equation}
where $\mu$ and $\lambda$ are two arbitrarily fixed real numbers.  The potential \eqref{1.2} is given by the selfadjoint extension of the operator $${\cal H}_0=-\frac{\hbar^2}{2m}\frac{d^2}{dy^2}$$ defined over $\mathbb{R}/\{0\}$ by the matching conditions
\begin{equation}
\left(
\begin{array}{c}
 \psi (0^+) \\ [1ex]
 \psi' (0^+) 
\end{array}
\right)=\left(
\begin{array}{cc}
 \displaystyle\frac{\hbar^2-m\lambda}{\hbar^2+m\lambda} & 0 \\  [2ex]
\displaystyle \frac{-2\hbar^2m\mu}{\hbar^4-m^2\lambda^2} & \displaystyle\ \frac{\hbar^2+m\lambda}{\hbar^2-m\lambda}
\end{array}
\right)
\left(
\begin{array}{c}
 \psi (0^-) \\  [1ex]
 \psi' (0^-)
\end{array}
\right),
\end{equation}
where we denote by $f(y_0^\pm)$ the limit of the function $f(y)$ when $y$ tends to $y_0$ from the left ($y_0^-$) and from the right ($y_0^+$).
This potential was studied in \cite{GNN} and has been shown to be relevant in physics. In fact,  Mu\~noz-Casta\~neda and Mateos-Guilarte \cite{JMG} had the idea to use a slight generalization of \eqref{1.2} given by
\begin{equation}\label{1.3}
V_{3}(y)=\mu_1 \delta(y+\ell/2)+\lambda_1 \delta'(y+\ell/2)+\mu_2 \delta(y-\ell/2)+\lambda_2 \delta'(y-\ell/2), 
\end{equation}
to mimic the physical properties of two infinitely thin plates with an orthogonal polarization interacting with a scalar quantum field. They evaluated the quantum vacuum interaction energy between the two plates and they found positive, negative, and zero Casimir energies depending on the zone in the space of couplings. This study was continued in \cite{gadella-jpa16}, finding new interesting results. For instance, when the limit $\ell\to 0$ is taken in \eqref{1.3}, the resulting potential supported at $x=0$ is a $\delta\text{-}\delta'$ potential with couplings given as functions of $\{\mu_1, \lambda_1,\mu_2,\lambda_2\}$, that defines a non-abelian superposition law.

All these results justify the use of a derivative of the delta  interaction along the delta itself. In the present paper, we propose the study of the properties of the one-dimensional periodic Hamiltonian 
\begin{equation}\label{1.4}
{\cal H}= -\frac{\hbar^2}{2m}\,\frac{d^2}{dy^2}+V(y), \quad \text{where} \quad
V(y)=\sum_{n=-\infty}^\infty [\mu\,\delta(y-ny_0)+\lambda\,\delta'(y-ny_0)],
\end{equation} 
where again $\mu$ and $\lambda$ are real numbers and $\ell>0$.  As usual, the time independent Schr\"odinger equation for this situation should be written as
\begin{equation}\label{1.5}
{\cal H}\psi(y)\equiv -\frac{\hbar^2}{2m}\,\frac{d^2}{dy^2}\,\psi(y)+V(y)\psi(y)={\cal E}\psi(y).
\end{equation}

In order to simplify expressions and calculations, it is usual working with dimensionless quantities. To this end, we use a redefinition of some magnitudes as was done in \cite{gadella-jpa16}. We perform this redefinition in three steps:
\begin{enumerate}
\item 
Magnitudes with the dimensions of a length are compared to the Compton wavelength:
\begin{equation}\label{1.6}
{\rm length\, magnitude}=\frac{\hbar}{mc}\cdot {\rm dimensionless\, magnitude} .
\end{equation}
This allows to introduce a dimensionless space coordinate over the line $x=y mc/\hbar$, as well as a dimensionless linear chain spacing $a=y_0 mc/\hbar$.

\item
The Dirac delta coupling $\mu$ has dimensions $[\mu]=ML^3T^{-2}$  and the Dirac $\delta^\prime$ coupling has dimensions $[\lambda]=ML^4T^{-2}$.  Hence we can introduce two dimensionless couplings $w_0$ and $w_1$ for the $\delta$ and the $\delta^\prime$ respectively, given by
\begin{equation}\label{1.7}
\mu = \frac{\hbar c}{2}  w_0,\qquad \lambda =\frac{\hbar^2}{m} w_1\,.
\end{equation}

\item
Energies are scaled in terms of $mc^2/2$,  so that from \eqref{1.4} the rescaled Hamiltonian is given by
\begin{equation}\label{1.8}
H= 2 {\cal H}/mc^2=-\frac{d^2}{d x^2}+\sum_{n\in \mathbb{Z}}\left[ w_0\delta(x-n a)+2w_1\delta^\prime(x-na)\right],
\end{equation}
and the eigenenergies of the time independent Schr\"odinger equation
\begin{equation}\label{1.777}
H\psi(x)=\varepsilon\psi(x)
\end{equation}
will be $\varepsilon=2 {\cal E}/mc^2$. 
\end{enumerate}
Notice that when rescaling the arguments of the $\delta$ and the $\delta'$ in \eqref{1.4} as
\begin{equation*}
{\delta(x-n y_0)=\frac{1}{y_0}\delta\left(\frac{x}{y_0}-n\right),\quad \delta'\left(x-ny_0\right)=\frac{1}{y_0^2}\delta'\left(\frac{x}{y_0}-n\right),}
\end{equation*}
 then the scale of energy defined by the strength of the $\delta'$ coupling is $\lambda/y_0^2$, and the scale of energy defined by the strength of the Dirac-$\delta$ is $\mu/y_0$. It is easy to see that the ratio between both energy scales defined by the strength of the couplings  can be written in terms of the Compton wavelength of the particle  $\lambda^C=\hbar/(mc)$ and the linear lattice spacing $y_0$ as
 \begin{equation}
\frac{\lambda}{\mu y_0}=\frac{2\lambda^C}{y_0}\frac{w_1}{w_0}.
\end{equation}
Typically, the lattice spacings in real crystals are of the order of  $y_0\sim \si{\angstrom}$ and the Compton wavelength for an electron is $\lambda^C_{e^-}=3.86\cdot10^{-3}\si{\angstrom}$. Hence  $\lambda/(\mu y_0)\sim 10^{-3}\frac{w_1}{w_0}$, meaning that $w_1$ should be much bigger than $w_0$ for the ratio of energy scales to be comparable. 
In the sequel, we shall always work with dimensionless quantities as defined above. 

The paper is organised as follows. In Section~\ref{sec_review} we review and generalise some basic results about the band structure for one-dimensional periodic potentials built from potentials with compact support smaller than the lattice spacing. In particular we remark some aspects that are not easily available in the standard  literature, such as the density of states. In Section~\ref{sec_1species} we present the original results we have obtained for a particular interesting example: what we will call on the sequel the {\it one-species hybrid Dirac comb}, which corresponds to the potential  \eqref{1.8} introduced before. Properties of the band spectrum, and density of states are studied in detail.
In Section~\ref{sec_2species} we deal with the {\it two-species hybrid Dirac comb}, obtained by adding an extra hybrid comb to  \eqref{1.8} displaced a distance $d$ with respect to the original. Finally in Section~\ref{sec_conclusions} we give our conclusions and further comments concerning our results.

\section{Review of band structure for one-dimensional periodic potentials}
\label{sec_review}

{In this section we present some general formulas relative to one-dimensional potential chains, with a periodic potential (built from a potential $V_C(x)$ with compact support $J_0=[-\eta/2,\eta/2]$) that vanishes outside small intervals $J_n=[na-\eta/2,na+\eta/2]$, included in $I_n=[na-a/2,na+a/2]$ centered around the chain points $na$ ($n\in {\mathbb Z}$ {and $a>\eta$}), whose union gives the whole real line. The dimensionless Schr\"odinger equation associated to $V_C(x)$ is:
\begin{equation}\label{2.1}
H_C \, \psi_k(x)\equiv \left(-\frac{d^2}{dx^2}+V_C(x)\right)\psi_k(x)=k^2 \psi_k(x),
\quad \varepsilon=k^2>0.
\end{equation}}
The potential $V_C(x)$ is not necessarily even with respect to spatial reflections $x\to -x$. For \eqref{2.1}, we find  two linearly independent scattering solutions: {one going from the left to the right ($R$) and the other in the opposite direction ($L$). 
Outside the interval $J_0$ these scattering waves have the following form}:
\begin{equation}\label{2.2}
\psi_{k,R}(x)=
\left\{ 
\begin{array}{ll}
 e^{-i k x} r_R(k)+e^{i k x}, & x<-\frac{\eta}2. \\ [2ex]
 t_{R}(k) e^{i k x}, & x>\frac{\eta}2.
\end{array}
\right.
\qquad 
\psi_{k,L}(x)=
\left\{ 
\begin{array}{ll}
t_{L} (k) e^{-i k x}, & x<-\frac{\eta}2. \\ [2ex]
e^{i k x} r_L(k)+e^{-i k x}, & x>\frac{\eta}2.
\end{array}
\right.
\end{equation}
The functions $\{r_R(k),r_L(k),t_R(k),t_L(k)\}$ represent right and left reflection and transmission scattering amplitudes. One interesting property of these coefficients is that $r_R(k)\neq r_L(k)$ if $V_C(x)\neq V_C(-x)$. On the other hand, time reversal symmetry of the Hamiltonian we are dealing with implies $t_R(k)=t_L(k)=t(k)$. The scattering matrix
\begin{equation}\label{2.3}
S=\left(\begin{array}{cc}
t(k) & r_R(k) \\[2ex]
r_L(k) & t(k)
\end{array} \right)
\end{equation}
is unitary. Therefore its two eigenvalues have modulus equal to $1$, and their respective arguments define the scattering phase shifts $\delta_\pm(k)$ in the  so called even ($+$) and odd ($-$) channels, respectively.

\noindent {Next, we construct the  periodic potential using $V_C(x)$ as building blocks. Thus, we have a Hamiltonian of the form:
\begin{equation}\label{2.4}
H_P=-{d^2}/{ dx^2}+V_P(x), \qquad \qquad V_P(x)=\sum_{n=-\infty}^\infty V_C(x-na).
\end{equation}}
In order to obtain the eigenfunctions of $H_P$, we need to use the Floquet-Bloch pseudo-periodicity conditions:
\begin{equation}\label{2.5}
\psi_q(x+a)=e^{i q a}\psi_q(x), \qquad 
\psi^\prime_q(x+a)=e^{i q a}\psi^\prime_q(x),\qquad  
q\in\left[-\frac{\pi}{a},\frac{\pi}{a}\right],
\end{equation}
where, as usual, we are restricting our considerations to the first Brillouin zone.
 
Since for each primitive cell $I_n$ the compact supported potential $V_C(x-na)$ vanishes outside the interval $ J_n=[n a- \eta/2,na+ \eta/2]$, 
then on any of the intervals $\mathcal{J}_n\equiv\{x\in I_n\vert\, x\notin J_n \}$ the Bloch waves are linear combinations of the two scattering solutions centered at the point $n a$:
\begin{equation}\label{2.6}
\psi_{k,n,q}(x)=A_n\psi_{k,R}(x-na)+B_n\psi_{k,L}(x-na)\,, \quad{\rm for}\,\,x\in 
\mathcal{J}_n\,.
\end{equation}
Then, we use the Floquet-Bloch pseudo-periodicity conditions given in equation \eqref{2.5} at the points $x=na-a/2$, 
so as to obtain the following two linear equations for the coefficients { $A_n$ and $B_n$:
\begin{equation}\label{2.9}
\left(\begin{array}{cc}
\psi_{k,R}(a/2)-e^{i q a}\psi_{k,R}(-a/2) & \psi_{k,L}(a/2)-e^{i q a}\psi_{k,L}(-a/2)\\[2ex]
\psi^\prime_{k,R}(a/2)-e^{i q a}\psi^\prime_{k,R}(-a/2) & \psi^\prime_{k,L}(a/2)-e^{i q a}\psi^\prime_{k,L}(-a/2)
\end{array}\right)\left(\begin{array}{c} A_n \\[2ex] B_n \end{array}\right)=0\,.
 \end{equation}  }
{Non trivial solutions in $A_n$ and $B_n$ for \eqref{2.9} only exist if the determinant of the square matrix in \eqref{2.9} vanishes. Using the scattering wave eigenfunctions \eqref{2.2},
we obtain the following secular equation \cite{NIE}:}
\begin{equation}\label{2.11}
\cos(q a)=\frac{e^{i a k} \left(t(k)^2-r_L(k) r_R(k)\right)+e^{-i a k}}{2\,
   t(k)}\,\equiv F(\varepsilon=k^2)
\end{equation}

Alternatively, taking into account that from the scattering matrix \eqref{2.3} 
\begin{equation}\label{tracedet}
{\rm tr}(S) = 2 t(k)\qquad \text{and}\qquad  \det (S)= t^2(k)-r_R(k)r_L(k), 
\end{equation} 
we can write the secular equation \eqref{2.11} as
{\begin{equation}\label{2.12}
{\rm tr}(S)\cos(q a)=e^{-i a k}+\det (S)e^{i a k}\,.
\end{equation} 
This equation enables to obtain the band energy spectrum as the different branches $\varepsilon= \varepsilon_n(q)$ of the function $F(\varepsilon)$ for $q\in[-\pi/a,\pi/a]$, that obviously, from \eqref{2.11}, are a symmetric function of $q$: $\varepsilon_n(-q)=\varepsilon_n(q)$. } Notice that for any branch $\varepsilon_n(q)$ taking the derivative of \eqref{2.11} with respect to $q$ we have 
\begin{equation}\label{2.16a}
-a \sin (q a)= \left. \frac{d F(\varepsilon)}{d\varepsilon}\right\vert_{\varepsilon_n} \frac {d\varepsilon_n}{dq }\,,
\end{equation}
and therefore, as the left hand side of \eqref{2.16a} only vanishes at $q=0,\pm\pi/a$ in the first Brillouin zone, we get the following important consequences:
\begin{itemize}
\item
The function $\varepsilon_n(q)$ is monotone on the intervals $(-\pi/a,0)$  and $(0,\pi/a)$. Otherwise it would have a critical point inside any of the intervals which would make $\frac {d\varepsilon_n}{dq }=0$, making the r.h.s of \eqref{2.16a} equal to zero. The function $\varepsilon_n(q)$ has either a maximum or a minimum at $q=0,\pm \pi/a$.
\item
As a consequence of the above, the function $F(\varepsilon)$ is monotone for all $\varepsilon$ in every branch $\varepsilon_n(q)$.

\end{itemize}
{Since the scattering amplitudes $\{t(k),r_R(k),r_L(k)\}$ have better analytical properties in the complex plane we can use equation \eqref{2.11} to write down the inequality that characterises the whole band spectrum} of the system in terms of either $k$ or the energy $\varepsilon=k^2$:
{\begin{equation}\label{2.17}
\left\vert  \frac{e^{i a k} \left(t(k)^2-r_L(k) r_R(k)\right)+e^{-i a k}}{2
   t(k)}\right\vert  
   = \left\vert  F(\varepsilon=k^2)\right\vert \leq 1\,.
\end{equation}
As shown in \cite{KURASOVLARSON}, the eigenfunctions at the band edges are of particular importance, i.e., those Bloch waves characterised by the values of the momenta $k_i$ such that 
\begin{equation}\label{2.18}
 \left\vert  F(\varepsilon=k_i^2)\right\vert= 1, \qquad i=0,1,2, \dots 
\end{equation}}
The discrete set of momenta satisfying \eqref{2.18} show the lower and higher value of $k$ for each allowed band. If for some of these points, say $k_i$, we have $k_i=k_{i+1}$ there is no gap between two consecutive bands. Furthermore, this is more probable to happen for high values of $k$ and then $|t(k)|$ is close to one (see Ref. \cite{KURASOVLARSON}).

There are two extreme situations. When the compact potential $V_C(x)$ is opaque, the transmission coefficient vanishes: $t(k)=0$. Under this conditions, equation \eqref{2.12} takes the form
\begin{equation}\label{2.19}
e^{-2 i k a}-r_R(k) r_L(k)=0\,.
\end{equation}
In this case the band equation becomes the secular equation of a square well with opaque edges, giving rise to a discrete energy spectrum. The other extreme situation is when the potential $V_C(x)$ is transparent: $ |t(k)|=1$. In this case, we do not have a band structure as from { \eqref{2.11}} the spectrum coincides with the free particle spectrum.

There is another possibility, which is the existence of negative energy bands (arising from localised states bound states of the potential with compact support from which the lattice is built, in case they exist). These are solutions of \eqref{2.11} for imaginary momenta, i.e., $k=i\kappa$, with $\kappa>0$, so that the energy is negative: $\varepsilon=-\kappa^2<0$. The allowed energies for {these bands} satisfy the following inequality:
\begin{equation}\label{2.21}
\left\vert  \frac{e^{- a \kappa} \left(t(i\kappa)^2-r_L(i\kappa) r_R(i\kappa)\right)+e^{ a \kappa}}{2
   t(i\kappa)}\right\vert\leq 1\,.
\end{equation}
Thus far, we have discussed the general form of the inequalities providing energy and momentum allowed bands for the periodic potentials under our consideration. Let us see now some properties of a crucial magnitude: the density of states.

\subsection{The density of states}\label{dos21}

The density of states $g(\varepsilon)$ in Solid State Physics contains the information about the distribution of energy levels. It plays a central role in the calculation of thermodynamic quantities from the physical properties defined by the quantum mechanical problem of one particle moving in the periodic potential that defines the crystal system, specially those magnitudes involving averages over {occupied levels} such as the internal energy, thermal and electric conductivity, etc.

This function $g(\varepsilon)$ is defined {as the} number of energy eigenvalues between $\varepsilon$ and $\varepsilon+d\varepsilon$ divided by the length of the first Brillouin zone $2\pi/a$.  {We may write the general expression for the density of states for a given band produced by a one-dimensional periodic potential as
\begin{equation}\label{2.22}
g_n(\varepsilon)= \frac{a}{2\pi} \int_{-\pi/a}^{\pi/a}  \delta(\varepsilon-\varepsilon_n(q))\, d q =
 \frac{a}{\pi} \int_{0}^{\pi/a}  \delta(\varepsilon-\varepsilon_n(q))\, d q,
\end{equation}
because $\varepsilon(q)=\varepsilon(-q)$. It is noteworthy that 
$\varepsilon(q)$ is multivalued and its $n$-th branch $\varepsilon_n(q)$ is the $n$-th energy band.
Equation \eqref{2.11} gives the energy as a function of the quasi-momentum $q$ for the $n$-th energy band, in terms of a function $\varepsilon_n=\varepsilon_n(q)$ for $q\in [-\pi/a,\pi/a]$.}
As already proven, the functions $\varepsilon_n(q)$ are monotone, and therefore if we make a change the variable $y_n=\varepsilon_n(q)$ we get
\begin{equation}\label{2.22b}
g_n(\varepsilon) = \frac{a}{\pi}  \int_{\varepsilon_n(0)}^{\varepsilon_n(\pi/a)} \frac{dq}{d y_n}\ \delta(\varepsilon-y_n)\, d y_n=
\frac{a}{\pi} \left\vert \frac{dq}{d \varepsilon} \right\vert  \int_{m_n}^{M_n} \delta(\varepsilon-y_n)\, d y_n,
\end{equation}
where $m_n= \min\{\varepsilon_n(0), \varepsilon_n(\pi/a)\}$ and $M_n= \max\{\varepsilon_n(0), \varepsilon_n(\pi/a)\}$. Then, the whole density of states will be
\begin{equation}\label{2.22c}
g(\varepsilon)=\sum_{n} g_n(\varepsilon) = \frac{a}{\pi} \left\vert \frac{dq}{d \varepsilon} \right\vert  \left(\sum_{n} \int_{m_n}^{M_n} \delta(\varepsilon-y_n)\, d y_n \right).
\end{equation}
Observe that the term in parentheses in \eqref{2.22c} is one if $\varepsilon$ belongs to any allowed band and is zero otherwise. From {\eqref{2.11}}, the explicit form of the function $q(\varepsilon)$ is
\begin{equation}\label{2.26}
q(\varepsilon)=\frac{1}{a} \arccos F(\varepsilon)\,.
\end{equation}
For the values of $\varepsilon$ outside any allowed band, the absolute value of the $F(\varepsilon)$ is bigger than one, and therefore $\arccos F(\varepsilon)$ becomes purely imaginary. This fact allows us to give a general expression for the density of states if we take into account that
\begin{equation}\label{2.28}
\left\vert \frac{dq}{d \varepsilon} \right\vert  \left(\sum_{n} \int_{m_n}^{M_n} \delta(\varepsilon-y_n)\, d y_n \right)
=
\left\vert  {\rm Re}\left(\frac{dq(\varepsilon)}{d\varepsilon}\right)\right\vert\,.
\end{equation}
Hence,
\begin{equation}\label{2.29}
g(\varepsilon)=\frac{1}{\pi}  \left\vert {\rm Re} \left[\frac{d}{d\varepsilon}\arccos F(\varepsilon) \right]\right\vert,
\end{equation}
a very important result that will be used in the sequel.

{If the charge carriers are fermionic particles, the probability of occupying a state of energy $\varepsilon$ is given by the  Fermi-Dirac distribution. Hence,  in order to obtain the average number of fermions per unit energy and unit volume, the density of states must be multiplied by the Fermi-Dirac distribution as follows:
\begin{equation}
N(\varepsilon)= g(\varepsilon) \frac{1}{e^{(\varepsilon-\mu)/T}+1}
\end{equation}
where $\mu$ is the chemical potential. $\mu(T)$ is also called the Fermi level and its value at $T=0$ is the Fermi energy. Zones with positive energy can turn out to be valence bands, and zones with negative energies, including the lowest one, can be conduction bands depending on the position of the Fermi energy. In addition, all the energies from the allowed zones belong to the absolutely continuous spectrum of the energy operator and the corresponding states are delocalized.}

On the other hand, if the charge carriers are bosonic particles the average number of particles per unit volume and energy follows the Bose-Einstein statistic:
\begin{equation}
N(\varepsilon)= g(\varepsilon) \frac{1}{e^{(\varepsilon-\mu)/T}-1}.
\end{equation}
It is noteworthy that at zero temperature, all bosons are localised in the minimum energy state giving rise under special circumstances to the Bose-Einstein condensation. Recently the system we are studying in this paper has been the focus of attention concerning the possibility of having Bose-Einstein condensation in one-dimensional periodic systems \cite{bordagJPA2019}.

\section{The one-species hybrid Dirac comb}\label{sec_1species}

In the present section, we discuss the periodic one-dimensional system with Hamiltonian given by $H_P$ as defined {in} \eqref{2.4}. We use the terminology of {\it  one-species hybrid Dirac comb} for this model. Hybrid because it combines the Dirac delta and its first derivative. The use of one-species will be clarified later when we introduce a two-species hybrid Dirac comb. Our objective is the determination and analysis of the band spectrum of $H_P$. In the previous section, we have seen that permitted and prohibited energy bands can be determined after inequalities like \eqref{2.17}, \eqref{2.18} and \eqref{2.21} that involve the modulus of the secular equation. As previously shown, this secular equation depends on the transmission and reflection coefficients for the scattering produced by a potential of the form $V_C(x)= w_0\,\delta(x)+2w_1\,\delta'(x)$, where $w_0$ and $w_1$ were given in Section 1.2. The explicit form of these coefficients were given in  \cite{GNN} and are
\begin{eqnarray}\label{3.1}
&\hspace{-0.8cm} \displaystyle t(k)=\frac{(1-w_1^2)k}{(1+w_1^2)k+i w_0/2}  , \quad & r_R(k)=-\frac{2 k w_1+i w_0/2}{(1+w_1^2)k+i w_0/2},  \quad
\displaystyle  r_L(k)=\frac{2 k w_1-i w_0/2}{(1+w_1^2)k+i w_0/2}, 
\label{3.1.5}
\end{eqnarray}
Then, replacement of \eqref{3.1} on \eqref{2.11}  gives
\begin{equation}\label{3.2}
{\cos( q a)= f(w_1) \left[ \cos (k a)+\frac{a}{2}w_0\ h(w_1)\,
 \frac{\sin (k a)}{k a}  \right]\equiv F(k;w_0,w_1)\,,}
\end{equation}
where the functions $f(w_1)$ and $h(w_1)$ are, respectively,
\begin{equation}\label{3.3}
f(w_1)=\frac{1+w_1^2}{1-w_1^2}  \,,\qquad
 h(w_1)= \frac{1}{1+w_1^2}\,.
\end{equation}
This result enables us to perform a detailed quantitative and qualitative study of the band spectrum and the density of states of the $\delta$-$\delta^\prime$ comb in the forthcoming subsections.

\paragraph{A brief remark on the generalised Dirac comb and the $\delta'$-potential} The one-species Dirac comb in Eq. \eqref{1.8} has been previously studied in Ref. \cite{refe1}. Nevertheless the definition used by the authors of the mentioned paper is not equivalent to the one used in this paper. Let us go into more detail to make clear the difference, and therefore clarify why different band spectra are to be expected in our case. To start with, the definition of $$\widehat K^{(1)}_{
0,w_1}=-\frac{d^2}{dx^2}+w_1\delta'(x)$$ shown in Ref. \cite{refe1} is given by a certain selfadjoint extension of the operator 
\begin{equation}\label{h0}
\widehat K=-\frac{d^2}{dx^2}
\end{equation}
acting on class-$(2,2)$ Sobolev functions over the space 
$$\mathbb{R}^*\equiv\mathbb{R}/\{0\}.$$ Specifically, the selfadjoint extension used to define the $\delta^\prime$ potential in Ref. \cite{refe1} is characterised by a domain of functions satisfying the matching conditions
\begin{equation}
\psi'(0^+)-\psi'(0^{-})=0, \,\,\psi(0^+)-\psi(0^-)=w_1\psi'(0),
\end{equation}
or equivalently,
\begin{equation}\label{dpr}
\left(
\begin{array}{c}
 \psi (0^+) \\
 \psi' (0^+) \\
\end{array}
\right)=\left(
\begin{array}{cc}
 1 & w_1 \\
 0 & 1 \\
\end{array}
\right)\cdot\left(
\begin{array}{c}
 \psi (0^-) \\
 \psi' (0^-) \\
\end{array}
\right).
\end{equation}
In contrast to Ref. \cite{refe1}, the Hamiltonian with a point interaction used in our manuscript
\begin{equation}
\widehat K^{(2)}=-\frac{d^2}{dx^2}+w_0\delta(x)+w_1\delta'(x)
\end{equation}
acting as well on the class-$(2,2)$ Sobolev space follows from \cite{GNN}, and is the selfadjoint extension of the operator \eqref{h0} characterised by the matching conditions
\begin{equation}
\left(
\begin{array}{c}
 \psi (0^+) \\
 \psi' (0^+) \\
\end{array}
\right)=\left(
\begin{array}{cc}
 \alpha & 0 \\
 \beta & \alpha^{-1} \\
\end{array}
\right)\cdot\left(
\begin{array}{c}
 \psi (0^-) \\
 \psi' (0^-) \\
\end{array}
\right),
\end{equation}
where
\begin{equation}
\alpha\equiv\frac{1+w_1}{1-w_1},\quad\beta\equiv\frac{w_0}{1-w_1^2}.
\end{equation}
It is straightforward to obtain the matching condition that characterises our definition of 
\begin{equation}
\widehat K^{(2)}_{0,w_1}=-\frac{d^2}{dx^2}+w_1\delta'(x)
\end{equation}
by just making $\beta=0$:
\begin{equation}\label{dpus}
\left(
\begin{array}{c}
 \psi (0^+) \\
 \psi' (0^+) \\
\end{array}
\right)=\left(
\begin{array}{cc}
 \alpha & 0 \\
 0 & \alpha^{-1} \\
\end{array}
\right)\cdot\left(
\begin{array}{c}
 \psi (0^-) \\
 \psi' (0^-) \\
\end{array}
\right).
\end{equation}
Clearly, comparing Eqs. \eqref{dpr} and \eqref{dpus} one can easily conclude that definition of the $\delta'$ is the different the as the one shown in Ref. \cite{refe1}. Furthermore, the definition of the $\delta'$ used in Ref. \cite{refe1} turns out to be {\it{\bf non local}}, meanwhile the one we used in our paper is {\it{\bf local}}. The comparison between both possible definitions has been discussed in Refs. \cite{ALB1,SE,FAS,m2,rpmF,fronphysG}.

\subsection{Analysis of the secular equation}

{From the general analysis carried out in the previous section the allowed energy gaps in this particular case are characterised by the condition $|F(k,w_0,w_1)|\leq 1$ where $F(k,w_0,w_1)$ is defined in \eqref{3.2}}. The solutions $\{k_i\}$, $i=0,1,2,3...$ of \eqref{2.18}  can also be characterised by the critical points of \eqref{3.2} in the sense that they are solutions of the equation
\begin{equation}\label{3.8}
\frac{d}{dk}\; \left[ \cos( k a)+\frac{a}{2}w_0\ h(w_1)\,
 \frac{\sin (k a)}{k a}  \right]=0\,.
\end{equation}
From \eqref{3.8}, we conclude that the limits between allowed and forbidden bands depend on the values of the parameters $w_0$ and $w_1$. 

Next, we proceed to the analysis of the band distribution in terms of the pair $(w_0,w_1)$. Here, we shall focus our attention on those values of $(w_0,w_1)$ that give rise to a limiting or critical behaviour. 

It is convenient to introduce a notation showing the dependence of the scattering coefficients with $w_0$, $w_1$ as well as with $k$. In the sequel, we shall write $t(k,w_0,w_1)$,  $r_R(k,w_0,w_1)$, $r_L(k,w_0,w_1)$ and $\delta(k,w_0,w_1)$, so that the dependence on the coefficients is manifested. In this context, we have singled out six cases:
\begin{enumerate}
\item
 {For $w_0=w_1=0$, we must have the free particle over the real line. This is indeed the case, since  $|t(k,0,0)|=1$ for any value of $k$ and, consequently, there is no band spectrum, but a complete continuum spectrum $\varepsilon\in(0,\infty)$. }

\item
{The case in which no $\delta'$ interaction is present ($w_1=0$, $|w_0|<\infty$), gives rise the standard Dirac delta one-dimensional comb \cite{kronig-prsa31}. }This implies that $f(0)=h(0)=1$, where $f(w_1)$ and $h(w_1)$ have been defined in \eqref{3.3} and that the secular equation \eqref{3.2} takes the well known expression
\begin{equation}\label{3.14}
\cos( q a)= \cos (k a)+ \frac{a}{2}\,w_0\, \frac{\sin (k a)}{ k a}\,.
\end{equation}

The band edge points are given by the discrete solutions, on $k_j$, of the following transcendental equations:
\begin{equation}\label{3.15}
\sqrt{\frac{4 k_j^2+w_0^2}{4 k_j^2}}\;\cos\left[k_ja+\arctan\left(\frac{2 k_j}{w_0}\right)-{\rm sg}(w_0)\frac{\pi}{2}\right]=\pm 1\,.
\end{equation}
Thus, we recover in this limiting case all the well known expressions for the Dirac comb.

\item
If $w_0=0$ with $w_1$ arbitrary but finite, there are no $\delta$-potentials, and there is only $\delta'$ potentials in the comb. For a pure $\delta^\prime$-potential all scattering amplitudes are independent of the energy. In fact,
\begin{equation}\label{3.11}
t(k,0,w_1)=\frac{1-w_1^2}{1+w_1^2}\,, \quad  r_R(k,0,w_1)=-\frac{2 w_1}{1+w_1^2} \,,\quad r_L(k,0,w_1)=\frac{2 w_1.}{1+w_1^2}\,.
\end{equation}

Moreover, the total phase shift vanishes:  $\delta(k,0,w_1)=0$. Hence the secular equation \eqref{3.2} simplifies to
\begin{equation}\label{3.13}
\cos( q a)=\frac{1+w_1^2}{ 1-w_1^2}\, \cos (k a)\,.
\end{equation}

\item
When $w_0$ is arbitrary, although finite, and $w_1=\pm 1$, the transmission coefficient vanishes, $t(k,w_0,\pm 1)=0$ and\footnote{It is of note that the critical values $w_1=\pm1$ occur when the parameter $\lambda$ in \eqref{1.7} satisfies $|\lambda|= \hbar^2/m=7.62\, {\rm eV}\si{\angstrom}^2$ for the electron. }
\begin{equation}\label{3.10}
r_R(k,\omega_0,-1)=r_L(k,w_0,1)=\frac{4 k-i w_0}{4 k + i\omega_0}\,, \qquad   r_R(k,w_0,1)=r_L(k,w_0,-1)=-1\,,
\end{equation}
therefore the potential is opaque at each node. Consequently,  as was shown in Ref. \cite{JMG}, when $w_1=+1$ the left edge ($x\to n a^-$) behaves {as a boundary with Dirichlet condition  and the right edge  ($x\to n a^+$) behaves as a boundary with Robin conditions}, and the opposite for $w_1=-1$. Hence, in the limit $w_1\to\pm 1$ the comb becomes an infinite collection {of boxes with length $a$ and} opaque walls where Dirichlet/Robin boundary conditions are satisfied in each side of the box. In this situation solutions to \eqref{3.2} are given by the discrete set $\{k_n\}$ satisfying the following transcendental equation
\begin{equation}\label{3.9}
\frac{\tan(k_n a)}{k_n a}=-\frac{4}{w_0 a}
\end{equation}

\item
Let us consider the limiting cases where $w_0=\pm\infty$ and $w_1$ is finite. In these cases, the transmission amplitude vanishes for all $k$, i.e., $|t(k,\pm\infty,w_1)|=0 $. There are solutions to the secular equation \eqref{3.2} only if $k_n=\frac{\pi}{a}n$, where $n$ is a positive integer. Hence the spectrum is purely discrete because all the allowed bands collapse to a point {($\varepsilon_n(q)$ becomes a flat line)}. The reflection amplitudes are constant, $r_R(k,\pm\infty,\omega_1)= r_L(k,\pm\infty,\omega_1)=-1$. Note that for a given integer $n$, there is an infinite number of eigenfunctions with energy $\varepsilon_n=k_n^2$ since the system becomes an infinite array of identical boxes with opaque Dirichlet walls.

\item 
When $w_1=\pm\infty$ the scattering data \eqref{3.1}-\eqref{3.1.5} gives $t(k)=-1$ and $r_R(k)=r_L(k)=\delta(k)=0$. Therefore in this situation we recover the free particle continuum spectrum.

\end{enumerate}

We finish the description of some relevant limiting cases at this point and we pass to study the band spectrum structure of the one-species hybrid Dirac comb.

\subsection{Structure of the band spectrum}

{So far, we have focused our attention on the positive band energy spectrum. As it is well known, the spectrum of positive energy bands contains an infinite number of allowed bands\footnote{This statement excludes the extreme situations in which one recovers the continuum spectrum of the free particle ($w_1=\pm\infty$ or $w_1=w_0=0$).}, since we are dealing with an infinite linear chain. Furthermore, when the coupling $w_0<0$, the $\delta\text{-}\delta'$ potential admits at most one bound state. This means that there might be situations in which the hybrid Dirac comb has one negative energy band (giving rise to non-propagating states) which can be characterised by pure imaginary momenta $k=i\kappa$ with $\kappa>0$.}  In the present subsection we study the conditions for the existence of a negative energy band and the conditions for the existence of a gap between the negative energy band (localised states) and the first positive energy band (propagating states).

\subsubsection{Negative energy bands in the pure $\delta$-comb}
To start with, we study these questions for the pure Dirac delta comb with attractive deltas. From \eqref{3.2}, it is easy to obtain the secular equation for the Dirac comb with attractive delta potentials by choosing $w_1=0$ and $w_0<0$, and purely imaginary momenta:
\begin{equation}\label{3.16}
\cos( qa) = F(i\kappa;w_0,0) \,, \qquad  F(i\kappa;w_0,0) =\cosh (\kappa a) - \frac a2\,|w_0|\,\frac{\sinh (\kappa a)}{\kappa a}\,.
\end{equation}

From Eq. \eqref{3.16}, we obtain the following relations:
\begin{equation}\label{3.17}
 F(0;w_0,0) =1-\frac a2\,|w_0| \qquad {\rm and} \qquad \left. \frac{\partial F}{\partial\,\kappa}(i\kappa; w_0,0) \right|_{\kappa=0} =0\,.
\end{equation}

After \eqref{3.17}, we observe that $\kappa=0$ is a critical point of the even function of $\kappa$ given by $F(i\kappa;w_0,0)$. In addition:

\begin{itemize}
\item
For the second derivative with respect to the variable $\kappa$, we have that
\begin{equation}\label{3.18}
\left. \frac{\partial^2 F}{\partial\,\kappa^2}(i\kappa; w_0,0) \right|_{\kappa=0} = a^2\left( 1- \frac{a}{6}\,|w_0|\right).
\end{equation}

\item
We have the following limit:
\begin{equation}\label{3.18p}
\lim_{\kappa \to \pm\infty} F(i\kappa;w_0,0)= +\infty\,.
\end{equation}

\item
Within the interval $0<|w_0|a<6$ and for $\kappa>0$, the first derivative of $F(i\kappa;w_0,0)$ is positive. This means that {$F(i\kappa;w_0,0)$} is also strictly monotonic.  Thus for $0<|w_0|a<6$, no relative extrema (maxima or minima) may exist, except for the minimum at the origin.
\end{itemize}
All these facts have important consequences that we list in the sequel:

\begin{itemize}
\item[$*$]
Values of $a$ and $w_0$ for which $0<|w_0|a<4$. There is a minimum of $F(i\kappa;w_0,0)$ at the origin with absolute value smaller than one. In consequence, the function $F(i\kappa;w_0,0)$ does not intersect the line $F=-1$ and intersects the line $F=1$ at some point  $\kappa_1$. Then, in principle, the valence band should be in the energy interval $[-\kappa^2_1,0] $, as these are the energy values for which the modulus of $F(i\kappa;w_0,0)$ is smaller than one. In additon, we must take into account the existence of the conduction band, which is characterised by the values of $k^2$ for which $|F(k;w_0,0)|\le 1$.  This gives an interval of energies $[0,k^2_2]$ for the conducting band. Therefore, we have a valence-conducting band in the energy interval $[-\kappa_1^2,k^2_2]$.

\item[$*$]
Values for which $4<|w_0|a<6$. From \eqref{3.18}, we see that $\kappa=0$ is still a minimum, although in this case, this minimum, $F(i0;w_0,0)$, is smaller than $-1$. In consequence, $\kappa^2=0$ is not an allowed value for the energy. The straight lines $F=\pm 1$ cut $F(i\kappa;w_0,0)$ at the points  $\kappa_1$ ( $F=1$)  and  $\kappa_2$ ($F=-1$). Since $F(i\kappa;w_0,0)$ is strictly growing on the semi-axis $\kappa\in(0,\infty)$, it comes out that $\kappa_2<\kappa_1$. Here, we have a valence band in the interval 
$[-\kappa_1^2,-\kappa_2^2]$.

\item[$*$]
Finally, we may have that $6<|w_0|a$. Now, $F(i\kappa;w_0,0)$ shows a {\it maximum} at $\kappa=0$ and a minimum at some point $\kappa_0$. The maximum, $F(i0;w_0,0)$ is smaller than $-1$ and so is the value of $F(i\kappa;w_0,0)$ at the minimum.  The situation is exactly as is in the previous case. 
\end{itemize}

\subsubsection{The band spectrum for the hybrid $\delta$-$\delta'$ comb}
We want to underline that the main objective of the present work is the analysis of the band structure differences  between the usual Dirac comb and the hybrid Dirac comb we have introduced in here. The strategy to accomplish this goal goes as follows:  first, we fix a value for $w_0$ and then study how the solutions of the secular equation (\ref{3.2}) vary with $w_1$. We do not expect to obtain an analytic closed form for this dependence. Instead, we rely on numerical and graphic methods with the aid of the package Mathematica.  

Let us go back to \eqref{3.2} and take $k=\sqrt\varepsilon$. We already know that the energy for the allowed bands satisfy the inequality
\begin{equation}\label{3.19}
\left\vert F(\sqrt{\varepsilon};  w_0,w_1)\right\vert=\left\vert f(w_1) \left[ \cos(a\sqrt{\varepsilon})+\frac{a}{2}w_0\ h(w_1)\,
 \frac{\sin(a\sqrt{\varepsilon})}{a\sqrt{\varepsilon}}  \right] \right\vert\leq 1\,.
\end{equation}
The choice of $\varepsilon$ as variable in \eqref{3.19} instead of $k$ has the purpose that just one single expression as \eqref{3.19} be valid for both {positive and negative energy bands}, depending on the sign of the energy $\varepsilon$. 
\begin{figure}[htbp]
\begin{center}
\includegraphics[height=0.22\textheight]{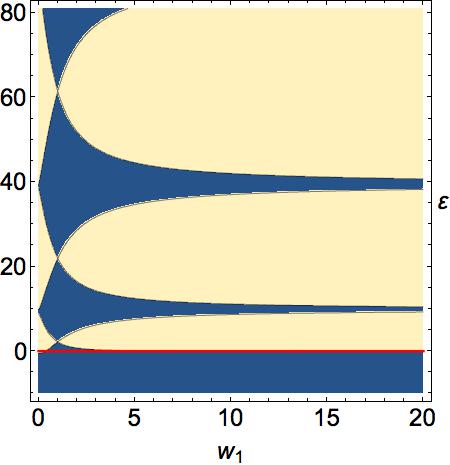}\qquad \qquad \includegraphics[height=0.22\textheight]{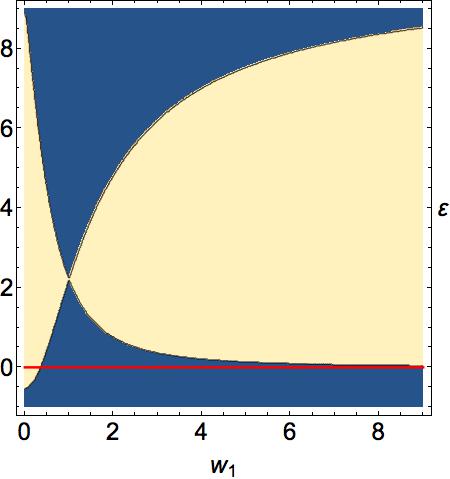}
\caption{(color online) On the left, allowed (yellow) and forbidden (blue) energy bands from \eqref{3.19} for $w_0=-0.5$ and $a=1$. On the right a zoom of the lowest energy band. The horizontal red line represents level $\varepsilon=0$.}
\label{Figure1}
\end{center}
\end{figure}

In Figures~\ref{Figure1}-\ref{Figure2}, we represent allowed ($\left\vert F(\sqrt{\varepsilon};  w_0,w_1)\right\vert\leq1$) and forbidden ($\left\vert F(\sqrt{\varepsilon};  w_0,w_1)\right\vert>1$) bands in terms of $w_1$ for given fixed values of $w_0$. The conclusions that we have reached after our results are the following:
\begin{figure}[htbp]
\begin{center}
\includegraphics[height=0.22\textheight]{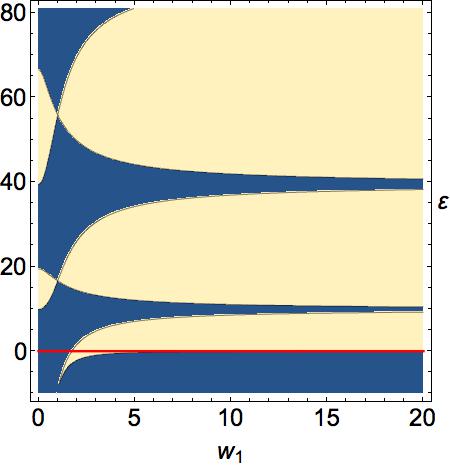}\qquad \qquad\includegraphics[height=0.22\textheight]{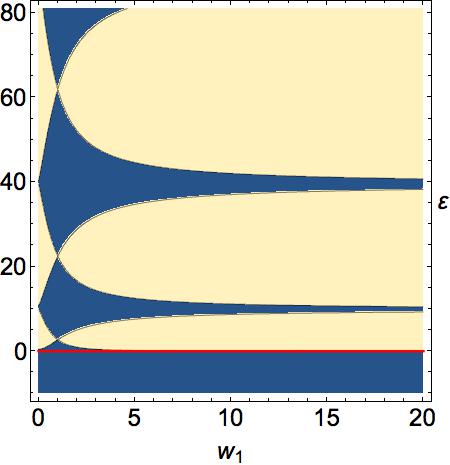}
\caption{(color online) On the left, allowed (yellow) and forbidden (blue) energy bands from \eqref{3.19} with $a=1$.Left: $w_0=-12$. Rihgt: $w_0=0.5$. The horizontal red line represents level $\varepsilon=0$.}
\label{Figure2}
\end{center}
\end{figure} 

\begin{enumerate}
\item
The opaque couplings $w_1=\pm1$ make the allowed energy bands collapse to isolated points, {so that we have a discrete energy spectrum coming from the well known secular equation \eqref{3.9}. This is in agreement with the numerical results shown in Figures~\ref{Figure1}-\ref{Figure2}: when $w_1=1$ the width of each allowed  energy band becomes zero}.

 \item
 For any given real value of $w_0$, the forbidden energy bands disappear in the asymptotic limit  $|w_1|\to \infty$, so that this limit gives the free particle.  The explanation of this apparently surprising outcome is the following: from \eqref{3.1} $t(k,w_0,\pm \infty)=-1$, thus the scattering due to the {$\delta-\delta'$ interaction} is almost transparent with just a difference of a phase shift $\pi$ after trespassing the $\delta'$ potential. Hence, when $|w_1|$ becomes large, the ``crystal'' effect disappears and the system behaves as the free particle on the real line. We reach the conclusion that the $\delta'$ coupling at $w_1=\pm\infty$ is not strong but on the contrary quite weak!
\end{enumerate}

 From the analysis of these plots we observe that there are regions in the  $(w_0,w_1)$-plane where there exists a negative energy band (the yellow regions below the horizontal red line in Figures \ref{Figure1} and \ref{Figure2}), and regions where there is no negative energy band, such as in the right plot of Figure \ref{Figure2} where the red line never intersects a yellow area. The situations in which the lowest energy band is positive correspond to a system where the charge carriers in the crystal move freely along it {as a plane wave}. Concerning the situation in which there is a negative energy band we can distinguish two very different behaviours\footnote{{The behavior of this system as a conductor or insulator, depends on the number of charge carriers in the crystal,  that together with the band spectrum fixes the position of the Fermi level.}}:
 \begin{itemize}
 \item 
 {When there is no gap between the negative energy band and the first positive energy band (regions where the red horizontal line is contained in the yellow area in Figures \ref{Figure1} and \ref{Figure2})  the carriers in the crystal can go from localised quantum states ($\varepsilon <0$) to propagating states ($\varepsilon>0$). This is a typical conductor behaviour when the charge carriers are fermions and the lowest energy band is not completely filled.
\item 
On the other hand when there is a gap between the negative energy band and the first positive energy band (regions where the red horizontal line is contained in the blue area in Figure \ref{Figure2}) all the carriers in the crystal are occupying localised quantum states ($\varepsilon<0$). The existence of such a gap demands an external energy input to promote carriers from the negative energy band to the positive one. Hence whenever the negative energy band is completely filled by carriers of spin $1/2$ this is a typical semiconductor or insulator behaviour depending on the size of the gap. }
\end{itemize}

\subsection{Dispersion relation and density of states in allowed bands} 

In our treatment of a particle moving through a periodic potential, it is noteworthy the emergence of some interesting features.

\subsubsection{Effect of the $\delta'$ on the energy bands}

Let us first consider the dispersion relation for each band $\varepsilon_n(q)$, given in \eqref{3.2} {making $k=\sqrt{\varepsilon}$. In Figures} \ref{Figure44}-\ref{Figure66} it is shown the behaviour of the energy band 
$\varepsilon_n(q;w_0,w_1)$ given by the solutions of the 
transcendental equation \eqref{3.2}. In each plot, the band $\varepsilon_n(q;w_0,w_1)$ is compared with the  corresponding energy band of the $\delta$ comb, which is $\varepsilon_n(q;w_0,0)$.
From Figures \ref{Figure44}-\ref{Figure66} we can infer the following general properties: 
\begin{figure}[htbp]
\begin{center}
\includegraphics[height=0.22\textheight]{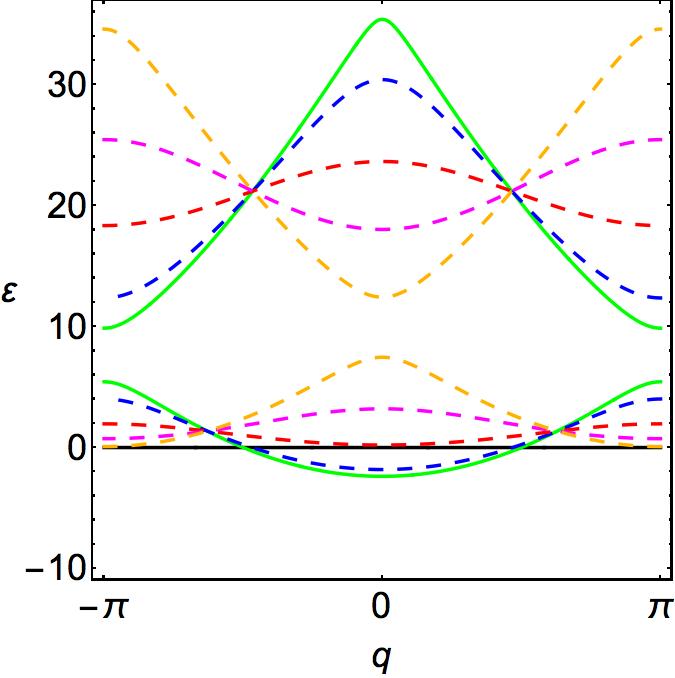}
\quad \quad
\includegraphics[height=0.22\textheight]{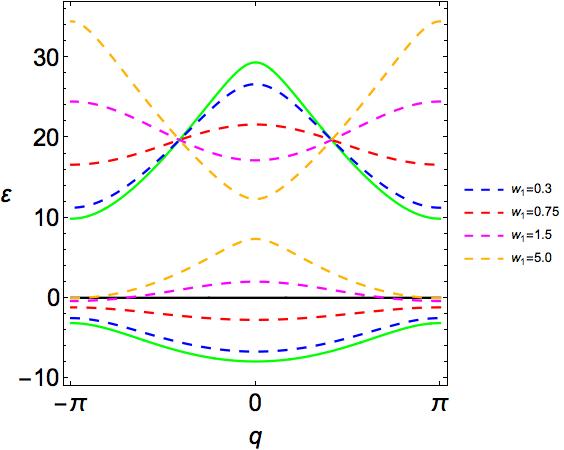}
\caption{(color online) First two allowed energy bands for the Dirac comb (solid green curve) and the one-species hybrid comb (dashed lines), given by \eqref{3.2}. 
On the left $w_0=-2$ for all the cases and on the right $w_0=-5$. In both cases the black line represents the zero energy level.}
\label{Figure44}
\end{center}
\end{figure}

\begin{enumerate}
\item
When the Dirac-$\delta$ comb has a completely negative energy band ($w_0<<0$), i.e. there is forbidden energy gap between localised states and lowest energy propagating states (see  Figure \ref{Figure44} right and \ref{Figure2} right), the appearance of a $\delta'$ term dramatically shifts the negative energy band towards higher energies.
Moreover, when the $\delta'$ coupling $w_1$ becomes large enough, {the negative energy band disappears to become a positive energy band. In addition, when the Dirac-$\delta$ comb is such that there is no gap between the highest negative energy state and the lowest positive one (see  Figure \ref{Figure44} left) the appearance of a $\delta'$ term does not shift the energy band towards higher energies.}

\item
When the Dirac deltas are repulsive (that is, $w_0>0$, see  Figure \ref{Figure66}), the appearance of a 
$\delta'$ term  shifts the maximum and minimum energy of each band. In any case $\varepsilon_n(qa=\pm\pi)$ decreases and $\varepsilon_n(0)$ {increases as $w_1$ does} for those energy bands $\varepsilon_n(q)$ such that $n=0,2,4...$. The effect is exactly the opposite for $n=1,3,5...$.
 Nevertheless, in this case the lowest energy band remains in any case within the positive energy region.

\item
In all cases, as can be seen on Figures \ref{Figure44}  and \ref{Figure66}, the introduction of the $\delta'$ term changes the curvature module of the allowed energy bands. On the one hand, whenever $|w_1|<1$ (subcritical values) the sign of the curvature of the allowed energy bands is the same as in the Dirac $\delta$ comb case. On the other hand, when $|w_1|>1$ (supercritical values) the sign of the curvature of the allowed energy bands changes with respect to the Dirac $\delta$ comb case. 

\begin{figure}[htbp]
\begin{center}
\includegraphics[height=0.22\textheight]{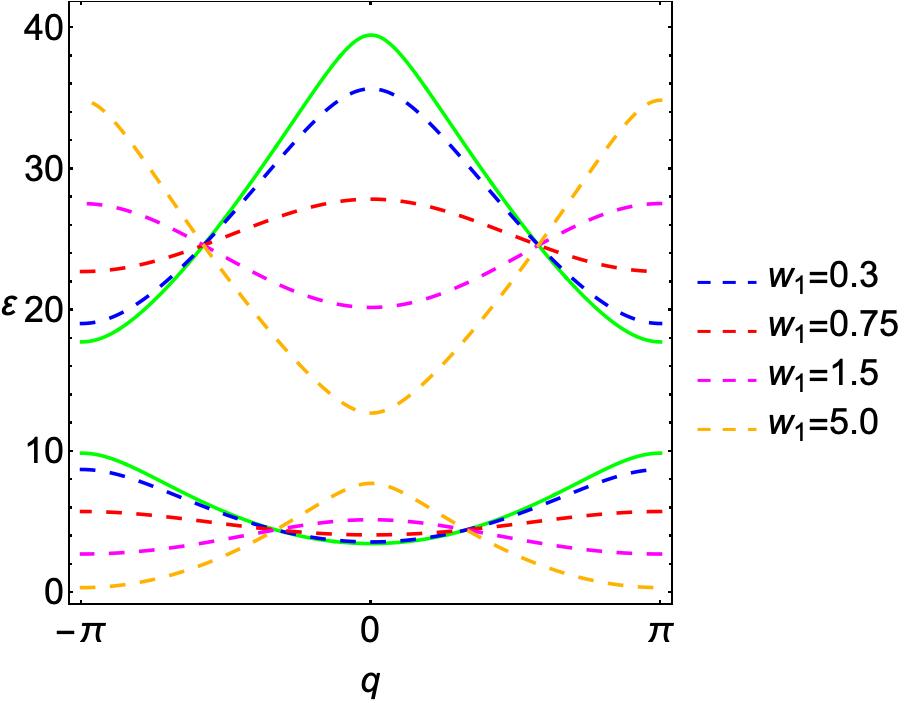}
\caption{(color online) First two allowed energy bands for the Dirac comb (solid green curve) and the one-species hybrid comb (dashed lines), given by \eqref{3.2}, with $w_0=5$ in all the cases. 
All the band have positive energy $\varepsilon$ because $w_0>0$.}
\label{Figure66}
\end{center}
\end{figure}

\item
From Figures \ref{Figure44}  and \ref{Figure66} it is straightforward to see that for fixed $w_0$, the $n$-th allowed energy band $\varepsilon_n(q;w_0,w_1)$ obtained for different values of $w_1$ have two fixed points that can be easily obtained from \eqref{3.2} and are given by
\begin{equation}
\frac{\tan (a\sqrt{\varepsilon})}{\sqrt{\varepsilon}}= -\frac4{w_0},
\end{equation}
{which correspond to each of the points of the discrete spectrum} obtained in \eqref{3.9} for the critical values $w_1=\pm 1$.

\end{enumerate}

\subsubsection{Effect of the $\delta'$ on the density of states}

\begin{figure}[h]
\begin{center}
\includegraphics[width=0.4\linewidth]{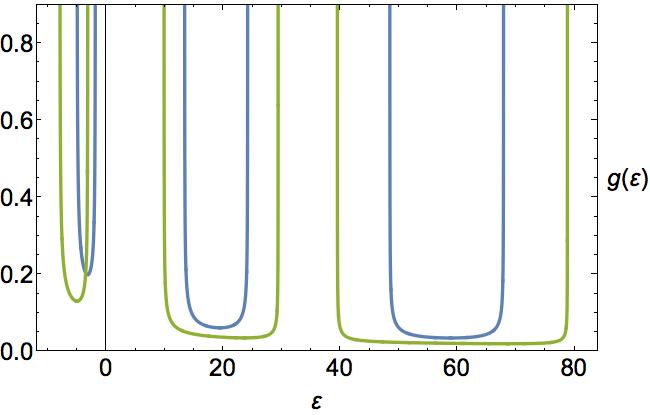}\qquad \includegraphics[width=0.4\linewidth]{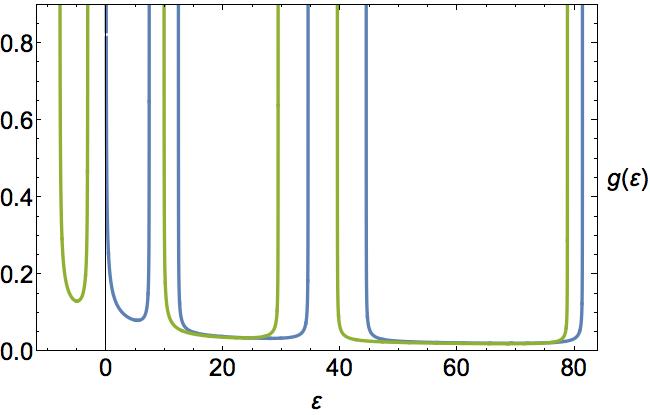}
\caption{(color online) Density of states in the lower bands of the $\delta$-$\delta^\prime$ comb. {On the left when $w_0=-5$ and  $w_1=0.5$ (blue curves), compared with the density of states in the Dirac comb (green curves). On the right analogous graphics when $w_1=5$.}}
\label{8}
\end{center}
\end{figure}

\begin{figure}[H]
\begin{center}
\includegraphics[width=0.4\linewidth]{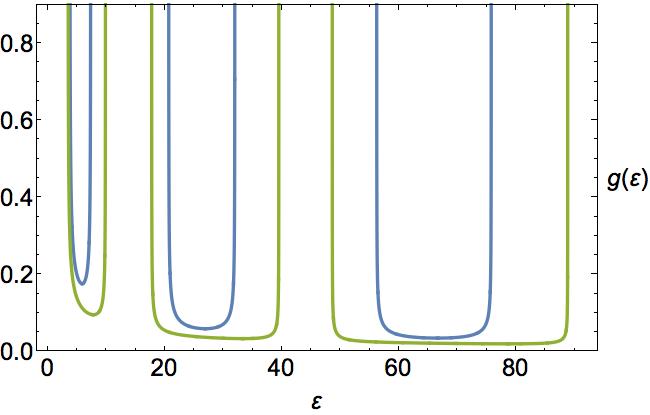}\qquad\includegraphics[width=0.4\linewidth]{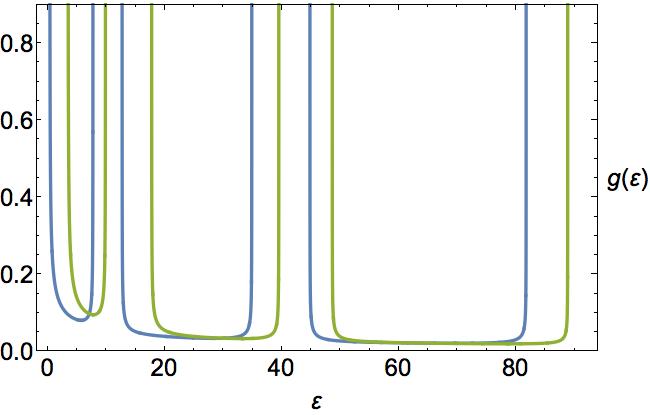}
\caption{(color online) Density of states in the lower bands of the $\delta$-$\delta^\prime$ comb. {On the left when $w_0=5$ and  $w_1=0.5$ (blue curves), compared with the density of states in the Dirac comb (green curves). On the right analogous graphics when $w_1=5$.}}
\label{9}
\end{center}
\end{figure}

The effect of the presence of a $\delta'$ on the density of states will be analysed next. Taking into account the results of Section~\ref{dos21}, and  in particular \eqref{2.29}, we show in
Figures~\ref{8} and~\ref{9} the properties of the density of states as a function of the energy $\varepsilon$ for the $\delta$-$\delta'$ comb in different situations. In addition, we compare the numerical results with the density of states of a Dirac $\delta$ comb with the same coupling.

In Figure~\ref{8} we show the typical behaviour of the density of states for strongly attractive delta wells with subcritical  and supercritical $\delta'$ couplings.
On the other hand, in Figure~\ref{9} we plot the density of states for strongly repulsive delta barriers with subcritical and supercritical  $\delta'$ couplings.

From the Figures, we infer the following general effects when we introduce a $\delta^\prime$ interaction in a Dirac comb:

\begin{itemize}
\item
Whenever $w_1$ is subcritical  ($|w_1|<1$), the band widths are narrower compared to the Dirac comb, and in addition the minima of the density of states is greater respect to the Dirac comb (see the left plots on Figures~\ref{8} and~\ref{9}).

\item
If $w_1$ is subcritical ($|w_1|<1$), the forbidden gap of the $\delta$-$\delta^\prime$ comb increases with respect to the Dirac comb (see the left plots on Figures~\ref{8} and~\ref{9}, and Figures \ref{Figure1}-\ref{Figure2} left as well).
If $w_1$ is supercritical ($|w_1|>1$), the forbidden gap of the $\delta$-$\delta^\prime$ comb decreases with respect to the maximum gap reached at the critical value $w_1=\pm1$  and tends to zero as $w_1\to\infty$ (see the right plots on Figures~\ref{8} and~\ref{9}, and Figures \ref{Figure1}-\ref{Figure2} left as well).

\item
For those cases in which the Dirac comb has a negative energy band ($w_0<0$), introducing the $\delta^\prime$ interaction shifts towards higher energies the lowest energy band (see Figure~\ref{8}). In addition, when $|w_1|>1$ the positive energy bands are as well shifted towards higher energies.
\item
When the lowest energy band of the Dirac comb is positive ($w_0>0$), introducing the $\delta^\prime$ interaction shifts towards lower energies the lowest energy band (see Figure~\ref{9}). This displacement of the energy happens as well for all the energy bands when $|w_1|>1$.

\end{itemize}

The qualitative effects just mentioned and shown in Figures~\ref{8} and~\ref{9} {are} maintained throughout the space of couplings $(w_0,w_1)$, a fact that can be inferred from Figures~\ref{Figure1}-\ref{Figure2} and other analytical studies of the densities of states and the forbidden energy bands \cite{KURASOVLARSON}.

\section{The two-species hybrid Dirac comb}\label{sec_2species}

The two-species hybrid Dirac comb is obtained by superposition of two one-species hybrid Dirac combs, like the potential in \eqref{1.8}, with different couplings and displaced by $\pm d/2$ with respect to the original one.
Therefore, in the two-species hybrid Dirac comb the potential $V_C(x)$ in \eqref{2.4} from which the periodic potential is built is given by
\begin{equation}\label{4.1}
V_C(x)=w_0\,\delta(x+\tfrac{d}2)+2w_1\,\delta^\prime (x+\tfrac{d}2)+v_0\,\delta(x-\tfrac{d}2)+2v_1\,\delta^\prime (x-\tfrac{d}2),
\end{equation}
which has been studied in detail in \cite{MM}. As it was already explained in Section~\ref{sec_review}, all we need to know to study the band spectrum and the density of states of the periodic potential built from \eqref{4.1} is its corresponding scattering data, which were computed in \cite{MM}:
 \begin{eqnarray}
 \label{4.3}
\!\! \!\! \!\! \!\! 
t(k)\!\!&\!=\!&\!\! \frac{1}{\Delta}\left(4 k^2 \left(v_1^2-1\right) \left(w_1^2-1\right)\right),
 \\ [1ex]
 \label{4.4}
\!\! \!\! \!\! \!\! 
r_R(k) \!\!&\!=\!&\!\! \frac{-1}{\Delta}\left( e^{-i d k} \left(2 k \left(v_1^2+1\right)+i v_0\right) \left(4 k w_1+i w_0\right)+e^{i d k} \left(2 k
   \left(w_1^2+1\right)-i w_0\right) \left(4 k v_1+i v_0\right) \right),
 \\ [1ex]
 \label{4.5}
\!\! \!\! \!\! \!\! 
r_L(k) \!\!&\!=\!&\!\! \frac{1}{\Delta}\left( e^{i d k} \left(2 k \left(v_1^2+1\right)-i v_0\right) \left(4 k w_1-i w_0\right)+e^{-i d k} \left(2 k
   \left(w_1^2+1\right)+i w_0\right) \left(4 k v_1-i v_0\right) \right),
 \\ [1ex]
\!\! \!\! \!\! \!\! 
\Delta(k)   \!\!&\!=\!&\!\! e^{2 i d k} \left(4 k v_1+i v_0\right) \left(4 k w_1-i w_0\right)+\left(2 k \left(v_1^2+1\right)+i v_0\right) \left(2 k  \left({w_1}^2+1\right)+i w_0\right).
 \end{eqnarray}
Our goal is to obtain the secular equation for this case. To this end, we operate as in Section~\ref{sec_review} for the single-species hybrid potential using the scattering data. Inserting these scattering amplitudes into \eqref{2.11} after some algebraic manipulations, and using the definitions introduced in \eqref{3.3} we obtain the following expression for the secular equation:
\begin{eqnarray}\label{4.8}
 \cos(q a)= f(w_1)f(v_1)\left[\frac{w_0 h(w_1)+v_0 h(v_1)}{2k} \sin (a k)+ \frac{v_0 w_1-v_1 w_0}{k} h(v_1) h(w_1) \sin (k (a-2 d))\right. \nonumber\\[2ex] 
+\left.  \left(1-\frac{v_0w_0}{4k^2} h(v_1) h(w_1)\right)\cos (a k)+ h(v_1) h(w_1)\left(4w_1v_1+\frac{w_0v_0}{4k^2}\right) \cos (k (a-2 d))\right]. 
\end{eqnarray}
Observe that there is only one term in \eqref{4.8} that breaks the exchange symmetry given by $(v_0,v_1)\leftrightarrow(w_0,w_1)$ and this is the coefficient of $\sin (k (a-2 d))$. Therefore, all those configurations of the two species comb
for which $v_0 w_1=v_1 w_0$
holds, are symmetric under the above exchange symmetry. In general, the band spectrum is symmetric under the following transformation
\begin{equation}\label{4.10}
(w_0,w_1,v_0,v_1,d)\leftrightarrow(v_0,v_1,w_0,w_1,a-d),\quad {0< d< a}\,.
\end{equation} 
This symmetry transformation is easily understood by recalling formula \eqref{4.8}. Indeed, the difference between looking at the $\delta\text{-}\delta'$ pairs at distance $d$ or to distance $a-d$ is the inversion of the roles of the coefficients $\{w_0,w_1\}$ and $\{v_0,v_1\}$.  This symmetry is shown by the secular equation \eqref{4.8}.

\subsection{The band spectrum for the two-species hybrid comb}\label{subsect41}
In this subsection we will carry out a qualitative study of the properties of the band spectrum for the two-species hybrid comb. In this situation, the space of parameters has dimension {5: $\{w_0,w_1,v_0,v_1,d\}$.  Looking} at equation  \eqref{4.8} the first thing we can infer is that there will be eight different possibilities of having a discrete spectrum, which correspond to regimes in the couplings in which the transmission amplitude \eqref{4.3} becomes {zero: the limits $w_1\to\pm1$,  $v_1\to\pm1$, $w_1=v_1\to\pm1$ and $w_1=-v_1\to\pm1$.}
\begin{figure}[htbp]
\centerline{ \includegraphics[height=0.22\textheight]{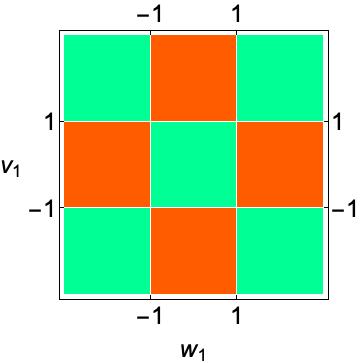}}
\caption{(color online) Regions in the $w_1$-$v_1$ plane that maintain the sign of the band curvature. The green areas are the ones in which each band of the hybrid comb has the same sign as the analogue band of the two-species $\delta$ comb. The orange areas are the ones in which each band of the hybrid comb has the opposite sign as the analogue band of the two-species $\delta$ comb.
}
\label{f9}
\end{figure}

As it happens for the one-species hybrid comb, whenever any of the $\delta'$-couplings {reaches} one of these critical regions of the whole coupling space, the bands of the comb become totally flat (zero curvature) giving rise to a pure point spectrum. Hence the critical regions mentioned in the items above are the regions where the sign of the curvature of the bands changes, with respect to the curvature of the bands of the pure two-$\delta$-species comb with couplings $w_0$ nd $v_0$. In Figure \ref{f9} we show the change of the band curvature for a hybrid comb with couplings $\{w_0,v_0,w_1,v_1\}$ with respect to the two-species $\delta$-comb with couplings $\{w_0,v_0\}$ in the $w_1$-$v_1$ plane.
The set of critical regimes described above can be divided into two sets:
\begin{enumerate}

\item{The four critical hyperplanes $w_1=\pm1$ and $v_1=\pm1$ affect only to one of the species. Thus, the real line is divided into independent boxes with opaque wall and length $a$. Each box confines a quantum particle on the interval $[na\pm d/2,(n+1)a\pm d/2]$, which consequently has a discrete set of energy values. These energy values are those obtained for an infinite one-dimensional square well with an additional interaction of the type $\delta\text{-}\delta'$ located at {$x=\pm d/2$. }The wave function satisfies Dirichlet boundary conditions at one side and Robin at the other. This fact was shown in  \cite{MM}.}

\item{The other four critical points correspond to the values $ w_1=v_1=\pm 1, w_1=-v_1=\pm 1$. In this case, all $\delta\text{-}\delta'$ interactions are opaque. Therefore, we have ``doubled'' the number of isolated boxes.  Each of the isolated intervals of length $[na\pm d/2,(n+1)a\pm d/2]$, which determine a box, is now split into two disjoint intervals:   
$$
[na\pm d/2,(n+1)a\pm d/2] \to [na-d/2,na+d/2)\cup(na+d/2,(n+1)a-d/2].
$$
At each wall, wave functions satisfy either Dirichlet, Neumann or Robin conditions as shown in \cite{MM}. }
\end{enumerate}

{The energy band spectrum is determined by the equation \eqref{4.8}.} In Figures \ref{f10}-\ref{f12} we show the first two energy bands for different two-species hybrid combs, compared to its analogue of two-species $\delta$-comb. From the figures we can infer the following general properties:
\begin{figure}[h]
\centerline{ \includegraphics[height=0.22\textheight]{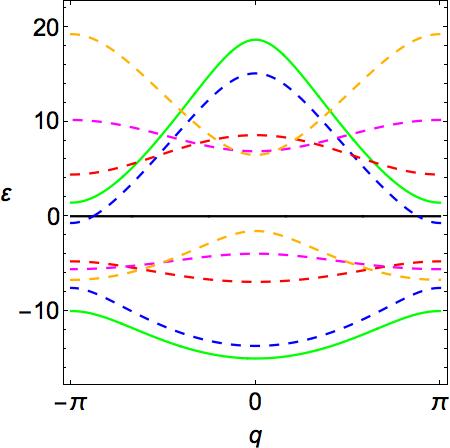}
\qquad\qquad
\includegraphics[height=0.22\textheight]{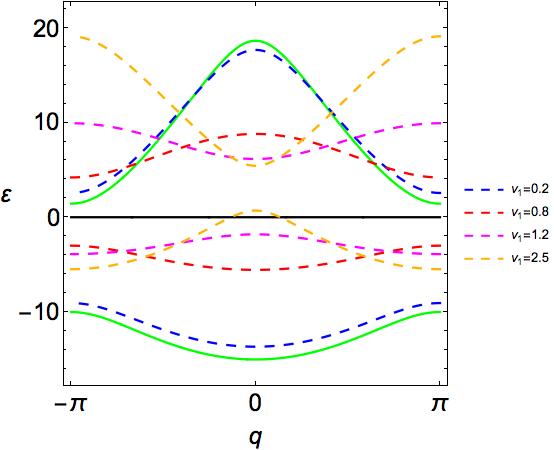}}
\caption{(color online) First two allowed energy bands for the two-species Dirac comb (solid green curve) and the two-species hybrid comb (dashed lines), given by \eqref{4.8}. For all the cases in both plots $w_0=-5$, $v_0=-6$, $d=1/3$, and $a=1$. Left: $w_1=0$. Right $w_1=0.2$.
}
\label{f10}
\end{figure}

\begin{figure}[h]
\centerline{ \includegraphics[height=0.22\textheight]{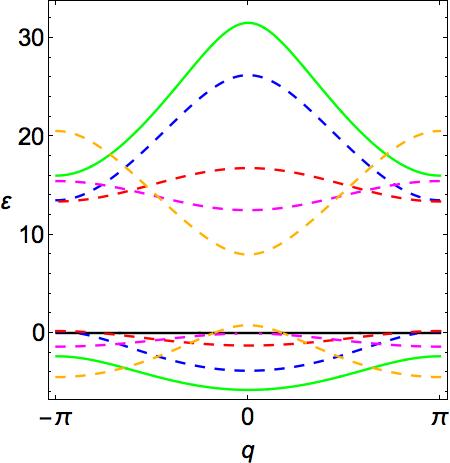}
\qquad\qquad
\includegraphics[height=0.22\textheight]{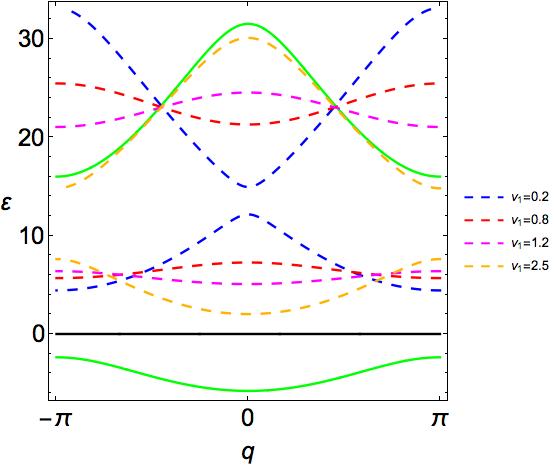}}
\caption{(color online)First two allowed energy bands for the two-species Dirac comb (solid green curve) and the two-species hybrid comb (dashed lines), given by \eqref{4.8}. For all the cases in both plots  $w_0=-5$, $v_0=5$, $d=1/3$, and $a=1$. Left: $w_1=0.2$. Right $w_1=4$.
}
\label{f11}
\end{figure}

\begin{figure}[h]
\centerline{ \includegraphics[height=0.22\textheight]{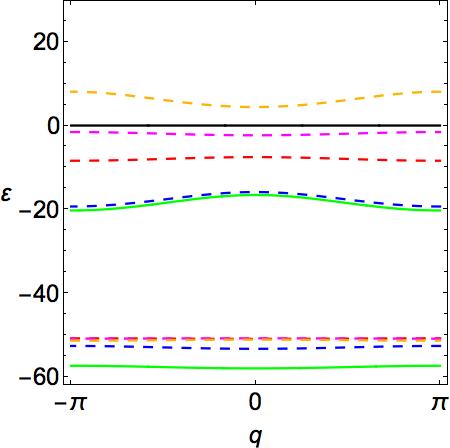}
\qquad\qquad
\includegraphics[height=0.22\textheight]{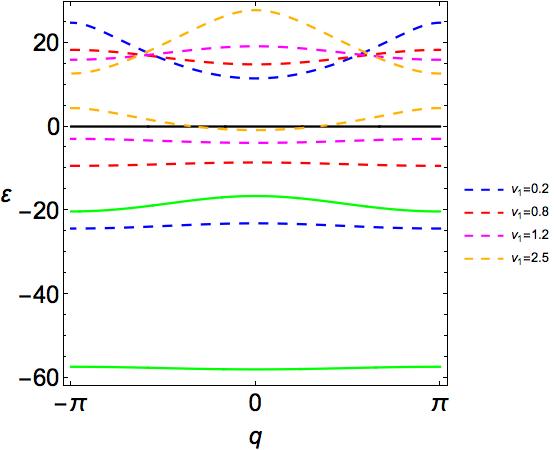}}
\caption{(color online) First two allowed energy bands for the two-species Dirac comb (solid green curve) and the two-species hybrid comb (dashed lines), given by \eqref{4.8}. For all the cases in both plots  $w_0=-15$, $v_0=-10$, $d=1/3$, and $a=1$. Left: $w_1=0.2$. Right $w_1=4$.}
\label{f12}
\end{figure}

\begin{itemize}
\item 
As can be seen from all the plots in Figures \ref{f10}-\ref{f12} a consequence of the existence of eight different possible discrete spectra is that there are none fixed crossing points for all the energy bands, unlike it happened in the one-species case for positive energy bands.

\item 
When both species of the hybrid comb include very attractive Dirac-$\delta$ wells (Figure \ref{f10}) and there is only one negative energy band, it mostly remains  in the negative energy part of the spectrum. Only in those cases in which at least one of the $\delta'$-couplings are supercritical, {$|w_1|\gg 1$ and/or $|v_1|\gg 1$, }the lowest energy band crosses to the positive energy spectrum (see Figure \ref{f10} right).

\item 
The energy shift produced by the appearance of $\delta'$ terms with respect to the two-species $\delta$-comb is much bigger for the negative energy bands and supercritical regimes. In fact as can be seen from all the plots in Figures \ref{f10}-\ref{f12} {right} this energy increase of the negative energy bands is such that they end up contained in a forbidden energy gap of the corresponding two-species $\delta$-comb as $w_1$ increases.

\item 
It is remarkable, that when the comb alternates a $\delta$-well and a $\delta$-barrier (i. e. $w_0$ and $v_0$ have opposite signs, as in Figure \ref{f11}), the lowest negative energy (localised states) band of the two-species $\delta$-comb becomes a positive energy band (propagating states) when one of the $\delta'$-couplings is in the supercritical regime, e. g. $|w_1|>>1$ (see Figure \ref{f11} {left}). 

\item 
The phenomenon described above happens as well for the excited negative energy band in those cases where there are two negative energy bands, as it is shown in the right plot of Figure \ref{f12}. 

\item 
Lastly it is quite interesting to remark, the physical properties of those hybrid combs with two negative energy bands (Figure \ref{f12}). The existence of regions in the space of parameters of the systems where one can find two negative energy bands is expected, since the double $\delta$-$\delta'$ potential admits two bound states, as it was shown in \cite{MM}. {These type of hybrid combs require very high temperatures to promote charge carriers from the lowest energy band to the first positive energy band, as can be seen from the right plot in Figure \ref{f12}. In fact an increase of temperature would promote the population of the excited negative energy band, provided that the crystal is not destroyed by such high temperature. Only in those cases in which the excited negative energy band becomes a positive energy band partially or totally, this first excitation would give rise to propagating states in the comb.}

\end{itemize}

\subsection{From two-species to one-species hybrid comb}

In this subsection we annalyse the limit in which the displacement of the combs $d$ tends to $0$ or $a$. This limit is of particular interest, because as it was shown in \cite{gadella-jpa16}, the superposition of two $\delta$-$\delta'$ potentials on the same point obeys a non abelian law.

To start with let us remember the basic result from \cite{gadella-jpa16}. {Given the potential \eqref{4.1} the limit $d\to 0$ gives rise to a single $\delta$-$\delta'$ potential
\begin{equation}\label{4.19}
\lim_{d\to0}V_{\delta\delta'}(x)=u_0\,\delta(x)+2u_1\,\delta^\prime (x),
\end{equation}}
where the couplings $u_0$ and $u_1$ are given in terms of the couplings $\{w_0,w_1,v_0,v_1\}$ by:
\begin{equation}\label{4.20}
u_0=\frac{v_0(1-w_1)^2+w_0(1+v_1)^2}{(1+v_1w_1)^2}
\,,\quad
u_1= \frac{v_1+w_1}{1+v_1w_1} 
\end{equation}
This result can be demonstrated by showing that the limit $d\to 0$ in the scattering data \eqref{4.3}-\eqref{4.5} results in the scattering data for a single $\delta$-$\delta'$ potential \eqref{3.1}-\eqref{3.1.5} with couplings $u_0$ and $u_1$ given by \eqref{4.20}.

\paragraph{The limit $d\to0$.} When we take the limit in which the displacement of the two-species hybrid comb 
tends to zero, taking into account the result given by \eqref{4.19} it is straightforward to see that we obtain a one-species hybrid comb with couplings $u_0$ and $u_1$ given by \eqref{4.20}. This case is a direct application of the result  obtained in \cite{gadella-jpa16}.

\paragraph{The limit $d\to a$.} {In this case, before using the central result from \cite{gadella-jpa16} we need to rearrange the comb appropriately. Notice that when $d$ gets close to $a$, then we can rewrite our original two-species hybrid comb  with the double $\delta$-$\delta'$ potential  \eqref{4.1} exchanging $w_0,w_1\leftrightarrow v_0,v_1$ centered at each linear chain point. Accounting for the traslation  invariance the result of taking $d\to a$ is a one species comb 
\begin{eqnarray}\label{4.22}
\sum_{n=-\infty}^{\infty}\tilde u_0\,\delta(x-na)+2\tilde u_1\,\delta^\prime (x-na)\,.
\end{eqnarray}}
where the resulting effective couplings are given by
\begin{equation}\label{4.23}
\tilde u_0=\frac{w_0(1-v_1)^2+v_0(1+w_1)^2}{(1+w_1v_1)^2}\,,\quad 
\tilde u_1=u_1= \frac{w_1+v_1}{1+w_1v_1}
\,.
\end{equation}
It is quite remarkable, that both limits give rise to a one-species hybrid comb. In both cases, the resulting $\delta'$-coupling is the same. Nevertheless, in each of the limits we obtain a different $\delta$-coupling

\section{Conclusions and further comments}\label{sec_conclusions}

{In this paper we have performed a detailed study of a generalised one-dimensional Kronig-Penney model using $\delta\text{-}\delta'$ potentials}. In Section \ref{sec_review} we have reviewed and generalised the formulas of the band spectrum and density of states for periodic potentials built from superposition of potentials with compact support centered in the linear lattice sites. As an application of the latter we have performed a very detailed study of the band spectrum for a hybrid comb formed by an infinite chain of identical and equally spaced $\delta$-$\delta'$ potentials .
It has been shown in previous works that the $\delta$-$\delta'$ potential
becomes opaque (identically zero transmission amplitude) when the coupling of the $\delta'$ satisfies $w_1=\pm 1$. Moreover,  it was demonstrated that when $w_1=\pm 1$ the two sides of the opaque $\delta$-$\delta'$ wall are equivalent to imposing Dirichlet (left-side)/Robin (right-side)  or Neuman (left-side)/Robin (left-side) boundary conditions. In both cases the Robin boundary condition parameter is determined by the Dirac-$\delta$ coupling $w_0$. As a consequence of this the most remarkable result concerning our study of the one-species hybrid comb is that the band spectrum degenerates to a standard discrete spectrum when we set $w_1=\pm1$. Moreover the addition of the $\delta'$ potentials with subcritical coupling ($\vert w_1\vert<1$) shows a narrower density of states distributions meaning that the density of states in the continuous spectrum is more concentrated
around the middle band point compared to the pure Dirac-$\delta$ comb. If the coupling of the $\delta'$ is supercritical ($\vert w_1\vert>1$) the width of the forbidden energy gaps decreases with $\vert w_1\vert$, reaching the free particle continuum spectrum for $\vert w_1\vert\to\infty$. To summarise, the effect of $\delta'$ interactions perturbing a Dirac-$\delta$ comb is more significant when we look at the curvature of the bands: while we remain in the subcritical regime $\vert w_1\vert<1$ the curvature of the bands stays the same as in the Dirac-$\delta$ comb, but crossing to the supercritical regime $\vert w_1\vert>1$ changes the curvature of the bands (see Figures \ref{Figure44} and \ref{Figure66}).

The conductor/insulator behaviour of the one-species hybrid comb requires a conceptual step forward to study the properties of the system with infinitely many charge carriers (see \cite{BMS1}). In addition when there are many charge carriers the spin-statistics properties must be accounted for. Our result on the calculation of the density of states enables to compute in future works the thermodynamical properties of these systems whenever the charge carriers are spin-$1/2$ particles, or integer spin particles (typically Copper pairs).

Lastly we have repeated the previous analysis for the two-species comb. This comb is built as an infinite chain of double $\delta$-$\delta'$ potentials. In addition to the appearance of eight opaque regimes in the space of couplings,
the allowed bands are deformed in interesting ways, even changing the curvature, with respect to the bands in the hybrid Dirac comb with only one-species of potentials. In this case the most remarkable effect over the curvature of the bands with respect {to} the one-species case, is that when both $\delta'$-couplings are in the supercritical regime, i. e.  $\vert w_1\vert,\vert v_1\vert>1$, the curvature of the bands remains the same as for the pure Dirac-$\delta$ two-species comb ($w_1=v_1=0$), as can be seen from the figures presented in Subsection~\ref{subsect41}.

\subsection{Further comments}
The connection between these type of dynamical systems and quantum wires was pointed and developed out by Cerver\'o and collaborators in Refs. \cite{KOR,KOR2}, where Dirac-$\delta$ chains are used a simple model for a quantum wire. Furthermore, when the Dirac-$\delta$ potentials are randomly distributed along the real line the authors found Anderson localization, and were able to reproduce  many of the physical properties expected in a quantum wire \cite{KOR2,KOR3,KOR4}. The authors were able to obtain further results on this system, such as a realistic absorption pattern in quantum wires when the coupling of the $\delta$-potentials is a complex number with positive imaginary part \cite{KOR3}, taking into account the previous results on ${\cal PT}$-symmetric periodic non-hermitian Hamiltonians \cite{KOR5,KOR6}. 

The basis of many of the works mentioned above is the fact that the Kronig-Penney comb is a very well studied periodic one-dimensional system. The results presented in this work generalise the Kronig-Penney comb, in order to provide a much richer model for quantum wires, where each point-supported potential in the chain contains two free parameters. The main physical consequence of introducing one extra coupling is that gives rise to a more tuneable band structure. In addition, the study carried out in this paper, where we have accounted for negative Dirac-$\delta$ couplings that give rise to negative energy bands, will enable to mimic absorption in the quantum wire when the system is studied in the quantum field theoretical framework at zero and finite temperature \cite{BMS1,BMS2}. In a quantum field theoretical framework, there is no need to assume complex couplings for the Dirac-$\delta$ to get the required unitarity loss. Furthermore, our results enable to extend the analysis performed for random distributions of Dirac-$\delta$ chains in the papers mentioned above, but for more general potentials with point support.

\section*{Acknowledgements}

This work was partially supported by the Spanish Junta de Castilla y Le\'on and FEDER projects (BU229P18 and VA137G18).  L.S.S. is grateful
to the Spanish Government for the FPU-fellowships
programme (FPU18/00957).
The authors acknowledge the fruitful discussions with M. Bordag, K. Kirsten, G. Fucci, and C. Romaniega.

\end{document}